# SWITCHABLE MOLECULAR FUNCTIONALIZATION OF AN STM TIP: FROM A YU-SHIBA-RUSINOV TIP TO A KONDO TIP

*Cosme. G. Ayani[1,2], Fabian Calleja[2], Ivan M. Ibarburu[1,2], Pablo Casado[1,2], Nana K. M. Nazriq[3], Toyo. K. Yamada[3,4], Manuela. Garnica[2], Amadeo L. Vázquez de Parga[1,2,5], Rodolfo Miranda[1,2,5]*

[1] *Departamento de Física de la Materia Condensada, Universidad Autónoma de Madrid, Cantoblanco 28049, Madrid, Spain.*
[2] *IMDEA-Nanociencia, Calle Faraday 9, Cantoblanco 28049, Madrid, Spain.*
[3] *Department of Material Science, Chiba University, 1-33 Yayoi-cho, Inage-ku, Chiba 263-8522, Japan.*
[4] *Molecular Chirality Research center, Chiba University, 1-33 Yayoi-cho, Inage-ku, Chiba 263-8522, Japan.*
[5] *Condensed Matter Physics Center (IFIMAC), Cantoblanco 28049, Madrid, Spain.*

## ABSTRACT

In this work we fabricate and characterize a functionalized superconducting (SC) Nb tip of a scanning tunnelling microscope (STM). The tip is functionalized with a Tetracyanoquinodimethane molecule (TCNQ) that accepts charge from the tip and develops a magnetic moment. As a consequence, in scanning tunnelling spectroscopy (STS), sharp, bias symmetric sub-gap states identified as Yu-Shiba-Rusinov (YSR) bound states appear against the featureless density of states of a metallic gr/Ir(111) sample. Although the coupling regime of the magnetic impurity with the SC tip depends on the initial absorption configuration of the molecule, the interaction strength between the superconducting tip and the charged TCNQ molecule can be reversibly controlled by tuning the tip-sample distance. The controlled transition from one coupling regime to the other allows us to verify the relation between the energy scales of the two competing many-body effects for the functionalized tip. Quenching the SC state of the Nb tip with a magnetic field switches abruptly from a tip dominated by the YSR bound states to a Kondo tip.

## INTRODUCTION

One of the main drawbacks of STM and STS is often the lack of detailed knowledge of the atomic and electronic structure of the tip of the microscope. In this sense STM tips were first funcitonalized with bulk materials in order to increase the energy resolution or to open the possibility to investigate the magnetic properties of surfaces [1-5]. For magnetic measurements it is crucial to control precisely the exact shape, "chemical identity" and magnetic state of the last few atoms at the apex of the tip and, in this sense, a controlled, well-characterized functionalization of the tip is of paramount importance. This is one of the reasons why in the recent years a great effort is being devoted to functionalizing STM tips with either nanoparticles, nanostructures or even molecules and adatoms. For example, CO-functionalized tips have increased the spatial resolution in both STM and non-contact Atomic Force Microscopy (AFM) [6], reversible functionalization of standard W tips with a nanosized superconducting cluster at the apex has increased the energy resolution [7], and functionalizing either a metallic tip with a magnetic molecule [8] or a superconducting tip with a magnetic atom [9] has provided new ways to probe locally the exchange field through the splitting of the Kondo resonance [8] or to quantify the spin polarization of magnetic samples through the changes in the intensity of the 100% spin polarized Yu-Shiba-Rusinov (YSR) sub-gap states [9]. The latter two alternative approaches to functionalize the tip, that rely in the development of either a Kondo resonance [8,10–17] or sub-gap YSR states [9,18–26], respectively, could be combined in a switchable single probe, as we describe in the present article.

The Kondo effect [8,10-17,27-31] predicts a many body ground state below a certain temperature, $T_K$, in which the spin of the magnetic impurity in a metal is screened by the antiferromagnetic interaction between the impurity spin and the conduction electrons of the metal host [29-31]. The Kondo effect has been extensively investigated by depositing on metal surfaces, magnetic adatoms [10], molecular radicals [13] (either organic or inorganic) or even non-magnetic molecules [12]. The hallmark of the Kondo effect in the local density of states (LDOS) is a zero-bias resonance (ZBR) with linewidth proportional to $2k_BT_K$ at zero temperature produced by the spin flip processes of between the conduction electrons and the magnetic impurity [13,30]. This ZBR can be directly probed by STS and, as Fano showed, the resulting STS spectrum can be explained as the interference produced by the two available tunnel paths for a tunnelling electron, one directly into the sample and the other indirectly via de ZBR [11,32].

Magnetic impurities on superconductors were theoretically predicted in the 1960s and first observed with STM in 1997 [21], the impurities produce an exchange scattering potential on the superconductor surface which misorients the spins in the Cooper pairs. In the limit of large impurity density, the localized spins of the impurities produce a gradual reduction of the superconducting gap with concentration. However, in the dilute limit, the pair-breaking potential produces bound states within the gap and localized at the impurity sites, called Yu-Shiba-Rusinov (YSR) bound states [24].

In the case of a spin ½ impurity, the YSR bound states give rise to a pair of sub-gap resonances symmetric in energy with respect to the Fermi level, easily detected by STS [22,33]. The sub-gap resonances reflect a quasiparticle excitation from the ground state to the first excited state via the particle or the hole component of the YSR states. The energy position of the sub-gap YSR resonances, $E_{BS}$, depends on the strength of the exchange interaction of the magnetic impurity, J:

$$E_{BS} = \Delta \frac{1-\alpha^2}{1+\alpha^2}$$

Where $\alpha = \pi \rho\, J\, S/2$, with ρ being the density of states at the Fermi level in the normal state, J the magnetic exchange and S the spin of the magnetic impurity. For weak coupling, i.e. negligible J (α ≈0), $E_{BS}$ ≈Δ and positive, while for strong coupling, large J, $E_{BS}$ ≈ -Δ (negative). Notice that there is quantum phase transition at $\alpha^2$ = 1, where $E_{BS}$ =0. The spectral weight of the particle- and hole- like excitation is influenced by the Coulomb potential (which breaks particle hole symmetry), and by asymmetries in the normal state conductance of the superconductor, although these contributions will be neglected in the following [24].

The strength of the exchange interaction determines in which of the two possible magnetic ground states the system is found: i) If the screening energy scale ($k_BT_K$) overcomes the pairing energy of the Cooper pairs (Δ), the impurity spin is completely Kondo screened , the ground state is a singlet (S=0) and a Kondo resonance is expected to be found outside the excitation gap and the system is said to be in the strong coupling regime [21]; ii) On the contrary, if $k_BT_K < \Delta$, the screening of the local spin is incomplete, the ground state has s=1/2 and the system is said to be in the free spin case or weak coupling regime [23,34,35]. Finally, a quantum phase transition from one to the other ground state occurs when $k_BT_K \sim \Delta$, and it is observed in tunnelling spectroscopy by the inversion of the relative spectral weight of the electron and hole components of the bound state as they cross the Fermi level [21,23,33,34].

As already stated these two many body effects, the Kondo resonance and the YSR states, result from the same exchange coupling strength, J, and, in fact, their energies are connected by a universal relation [26,27]. In principle, one should expect to observe the transition from one effect to the other by quenching the superconducting state of the host either by means of temperature or of magnetic field. In this way the competition between both phenomena may be studied [16,19,20] corroborating for different systems the predicted relation between their energy scales [27,28].

In this work, we manipulate vertically a purely organic and, initially, non-magnetic Tetracyanoquinodimethane molecule (TCNQ) from a graphene/Ir(111) substrate onto a superconducting Nb STM tip. Upon adsorption on the STM tip, the molecule charges and develops a magnetic moment which produces two bias-symmetric intra gap states in STS that are assigned to a pair of YSR bound states. The initial interaction strength between the magnetic impurity and the SC tip is determined by the adsorption configuration of the molecule onto the tip apex. We are able, however, to change continuously the strength of the exchange interaction, J, by controlling the distance between the functionalized-tip and the sample. This allows us to explore the quantum phase transition between two different magnetic ground states (free spin and Kondo screened, respectively) related to the weak ($k_BT_K < \Delta$) and strong ($k_BT_K > \Delta$) coupling of the magnetic impurity to the SC tip. We also verify the relation between the energy scale of the two many body effects, $E_{BS}$ (energy position of larger bound state) and $k_B T_K$, respectively, showing that the results obtained with our functionalized tips are consistent with the universal relation predicted by Matsuura [27,28] that connects the relative strength of both spin-dependent scattering processes. Finally, we demonstrate the *in–situ* reversible switching from a YSR tip to a Kondo tip by quenching the superconductivity of the Nb tip with an external magnetic field.

## **RESULTS & DISCUSSION**

The SC tips used to perform these experiments were produced by electrochemically etching a Nb wire (see *supplementary information*). The resulting STM tips have an average gap at 1.2 K of $\Delta = 1.05\ mV$, Fig. S1(a), and are fully capable of attaining atomic resolution as shown in Fig. S1(b) where both the atomic periodicity and the standing waves of Cu (111) are resolved simultaneously. The tips have been characterized by their temperature and magnetic field dependence yielding a critical temperature and critical magnetic field of 8.4 K and 1.2 T (Fig. S1(c) and (d)), respectively. These values differ from their bulk counterparts, especially in the case of the critical magnetic field due to tip size and geometry [35-37].

TCNQ molecules were deposited onto graphene/Ir(111) (hereafter called for short gr/Ir). The two distinctive characteristics of gr/Ir are the preservation of the Dirac cone of graphene with a slight p-doping that leaves the Fermi level around 110 meV below the Dirac point [38,39-41], and the incommensurate (9.3x9.3) moiré pattern produced by the mismatch between the two atomic lattices [42]. Once the molecules are absorbed, as reported previously, the graphene sheet effectively decouples the TCNQ molecules from the metal underneath [43,44]. The presence of the gr/Ir moiré pattern does not affect the molecule organization, as they show no preferential absorption sites within the moiré unit cell. The molecules form close packed, well-organized islands which are stabilized by hydrogen bonds between the neighbouring molecules and are bonded to the surface by π- π* interaction [45]. Although these molecules are well-known electron acceptors [45,46], STS measurements reveal a negligible charge transfer from the gr/Ir(111) substrate to the molecules, consistent with the above mentioned slight p-doping of the graphene layer. Finally, the molecular coverage of the surface is kept intentionally below the monolayer, as shown in Fig. 1(a), for two reasons. First, to facilitate the vertical manipulation of the molecular species onto the apex of the SC Nb tip. Second, to use the patches of gr/Ir not covered with TCNQ molecules to perform all STS measurements, as explained in the following.

In order to manipulate vertically a molecule onto the tip apex and functionalize the Nb tip, the edge of the molecular islands was scanned at high speed between 75 and 110 nm/s with currents above 1 nA while the sample was kept at 1.2 K. In these conditions, with our typical feedback loop bandwidth of the order of 200 Hz, the interaction between the tip and the molecules at the edge of the island is large enough to produce the tip functionalization. When a change in relative tip-sample distance is observed, the scanning is stopped and the tip is moved to a TCNQ-free area of the gr/Ir(111) substrate, where STS is performed (see section S2 in SI).

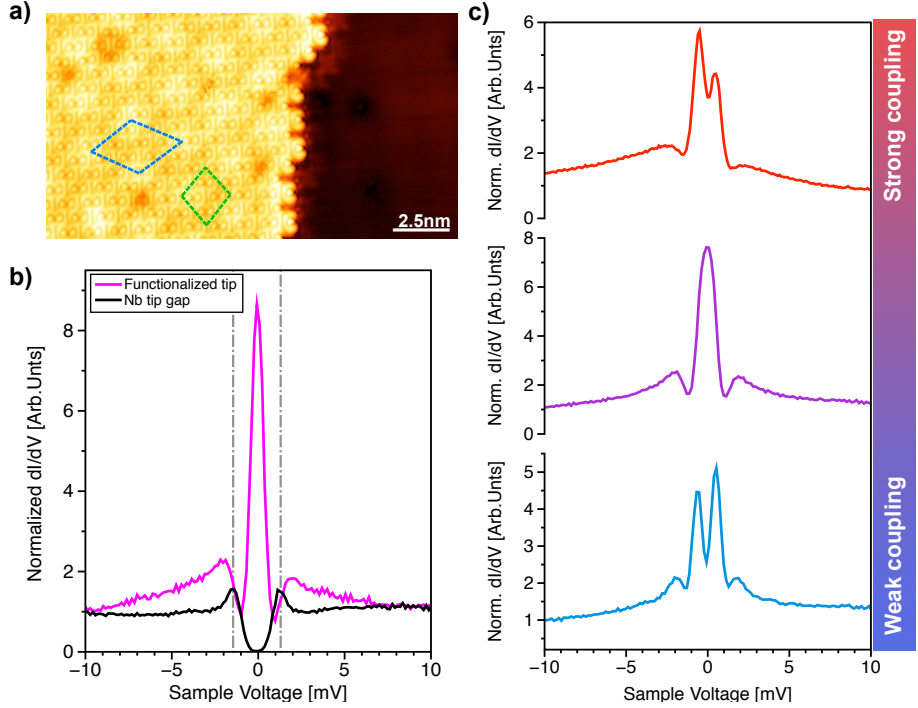

**Figure 1.** *a) STM topographic image (V=2 V, I=0.1 nA, T=1.2 K) showing the surface of gr/Ir(111) partially covered with TCNQ molecules. The moiré pattern periodicity of the substrate (blue dashed rhombus) can be resolved through the TCNQ layer. The molecular unit cell is indicated by the green dotted line. b) Differential conductance spectra recorded on a clean area of gr/Ir(111) before (black) and after (violet) manipulation of a TCNQ molecule onto the apex of the Nb STM tip. Stabilization parameters V=15 mV, I=0.5 nA, $V_{mod}$= 150 µV and T=1.2 K. c) dI/dV spectra of three representative TCNQ-functionalized tips. As indicated by the colour code at the right of the image and the asymmetry of the peaks, they represent the two different coupling regimes and the quantum phase transition discussed in the text. STS parameters: V=15 mV, I=0.5 nA, $V_{mod}$=100 µV and T=1.2 K.*

While the *dI/dV* spectrum recorded with the clean tip only shows its superconducting gap, the molecular functionalization of the tip results in the appearance of sub-gap states (Fig. 1(b)). The process is reversible. If the functionalized tip is pulsed at +2 V on the clean gr/Ir patches (bias voltage applied to the sample), the TCNQ molecule is released from the tip apex and the SC gap is recovered in the tunnelling spectrum (see Fig. S2(a) and (b)). The manipulation procedure is completely reproducible: every tip successfully functionalized with a TCNQ molecule shows sub-gap states. Such states may vary slightly in relative intensity or position inside the SC gap, but all of them can be classified into the three types of spectra depicted in Fig. 1(c). All spectra are taken against clean, metallic gr/Ir surfaces with a constant LDOS in the region explored which, accordingly, do not contribute to the recorded spectra.

In Fig. 1(c), we show the STS spectra of different TCNQ-functionalized tips. They resemble the shape expected for the YSR bound states due to a spin ½ impurity on a superconducting substrate, with asymmetric intensities for the electron-like and hole-like resonances at positive and negative voltages, respectively. As it will be confirmed by the results presented further up on the article, we can make the following assignments: The blue spectrum corresponds to the STS of a magnetic impurity weakly coupled to a superconductor, with the magnetic impurity in a free spin ground state (S=1/2), the superconductor ground state fully paired and the sub-gap state unoccupied, ($E_{BS}$ >0) [22,24,37]. The red spectrum corresponds to the one of a magnetic impurity in the strong coupling (or completely Kondo screened) regime, with the superconductor ground state involving an unpaired electron screening the magnetic impurity (S=0) and the bound state occupied ($E_{BS}$ <0) [22,24,37]. The spectrum in the center (purple) with a peak at zero energy would correspond to a magnetic impurity in the quantum phase transition between the two coupling regimes [22,24,37] that occur at a critical strength of the exchange coupling, $J_{cr}$, where the ground and excited state reverse their role.

TCNQ is a strong electron acceptor, well known to accept charge transferred from metal surfaces with low work function [46] or even from donor molecular species [47]. Although a TCNQ molecule physisorbed on the surface of gr/Ir(111) shows no charging of the lowest unoccupied molecular orbital (LUMO) [45], we propose that the TCNQ acquires charge from the superconducting Nb tip once it is manipulated onto its apex. Furthermore, it is known that individual TCNQ molecules that have accepted enough charge from the substrate as to half fill their LUMO orbital can acquire a substantial net magnetic moment [48]. As a consequence, a Kondo resonance appeared at zero bias in the STS measurements with normal tips [12, 47, 48].

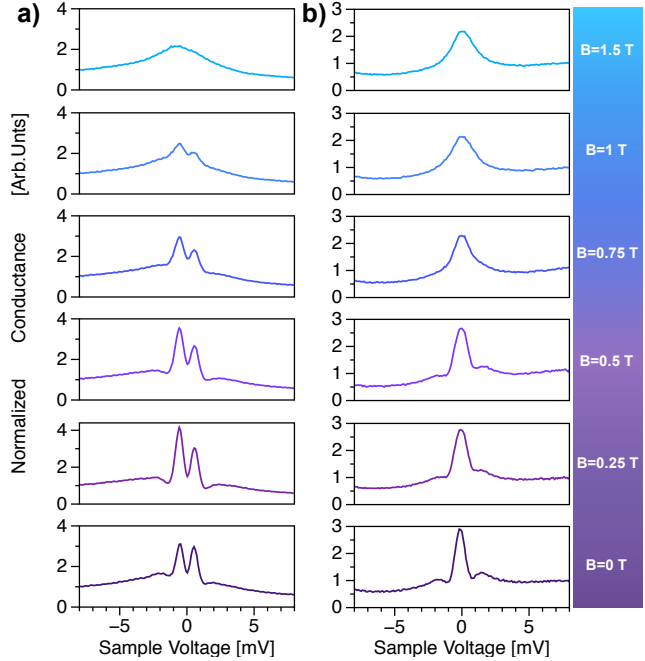

**Figure 2**. *Two sets of experiments performed with different TCNQ-functionalized Nb tips on bare gr/Ir(111) substrates. The magnetic field is ramped above the critical field of the Nb tip (1.2 T) to quench its SC state. The quenching unveils the Kondo resonance at the Fermi level demonstrating that the TCNQ molecules have acquired a magnetic moment. **a)** Evolution in the case of a TCNQ molecule in the strong coupling regime ($E_{BS}$< 0). **b)** In this case the molecule is initially in the quantum phase transition ($E_{BS}$≈0). Stabilization parameters are V=15 mV, I=0.5 nA, V $_{mod}$= 150 µV and T=1.2 K.*

In order to demonstrate that the TCNQ molecule transferred to the Nb tip has acquired a net magnetic moment and, thus, behaves as a magnetic impurity, we quench the SC state of the Nb tip by applying a magnetic field perpendicular to the sample surface (and therefore parallel to the tip axis). Figure 2 shows the result for two tips functionalized with TCNQ molecules in different coupling regimes. In both cases, when the magnetic field is increased above the critical field of the SC tip (B=1.5 T), a broader Zero Bias Peak (ZBP) remains indeed at zero energy.

To prove that these zero bias resonances observed after quenching the SC state are indeed Kondo resonances, we performed temperature dependent experiments for two functionalized tips whose molecules are in the strong coupling regime and quantum phase transition, respectively. Fig. S5 shows the dependence of the intrinsic full width at half maximum (FWHM) with temperature for the Kondo resonances of the two functionalized tips. In both cases the sets of data follow the expected temperature dependence for a Kondo resonance in the Fermi liquid theory with free $\alpha$ fit [11,13,14]. The results are in good agreement with those of the magnetic field experiments as the Kondo temperature of the molecule in the strong coupling regime (23.8 K) is higher than that of the molecule in the quantum phase transition (10.8 K). These results confirm that the molecule charges up and develops a magnetic moment when vertically manipulated from the surface of gr/Ir onto the SC Nb tip apex.

In order to quantify the strength of the Kondo coupling, the corresponding spectra obtained after quenching the superconductivity of the tip with a magnetic field at 1.2 K (Fig.2) are fitted with the Fano function taking into account both thermal and modulation broadenings, and the obtained intrinsic FWHM is then corrected by subtracting the corresponding Zeeman splitting for the case of a spin ½ magnetic impurity (see Fig. S4 of supplementary material) [49]. As it may already be observed in Fig. 2, the Kondo resonance produced by the molecule in the strong coupling regime has a larger intrinsic FWHM (5.7 meV) than the TCNQ molecule in the quantum phase transition (2.1 meV).

Likewise, the energy of the magnetically induced bound YSR states ($E_{BS}$) can be obtained by fitting the dI/dV spectra with the Cauchy function [33]. For a molecule in the strong coupling regime, the larger bound state is below the Fermi level ($E_{BS}<0$) while for a TCNQ molecule in the weak coupling regime it is above the Fermi level ($E_{BS}>0$). For the same functionalized tips (same molecule and configuration), the energy scale of the Kondo screening channel ($k_B T_K$) can be obtained from the Fano fit of the spectra upon quenching the SC state of the tip. As a one to one correspondence has been stablished between the position of the YSR bound states and the linewidth of the Kondo resonance for a same molecule, the competition between both many body effects can be studied by plotting the energy value of the magnetically induced bound states against the energy of the screening channel ($E_{BS}/\Delta$ vs $k_B T_K/\Delta$)) as shown in Fig. S6 [22,25,26] and compare it to the universal relation introduced by Matsuura [27,28].

Matsuura's universal relation describes the competition between the Kondo screening and the induced YSR bound states. When a magnetic impurity is placed on a superconductor the spin of the magnetic impurity becomes under screened as there is a decrease in the number of itinerant electrons that can participate in the Kondo cloud. Matsuura predicted that the weaker the Kondo screening the more strongly bound will be the sub-gap states should be. If the energy value of the sub-gap states ($E_{YSR}$) is plotted against the energy of the Kondo screening channel ($k_B T_K$) the data should follow the relation predicted by Matsuura (green dashed line show on Fig.S6) [27,28]. Therefore, it would be very interesting to develop a method to modify the coupling regime of the magnetic impurity, in order to access whole spectrum of the interaction strength of a magnetic impurity.

We have no control over the coupling regime established when the molecule is manipulated onto the tip apex, as it depends on the initial adsorption configuration of the molecule on the tip. Statistical analysis of 15 different functionalized tips show that it is more probable that the molecule ends in the quantum phase transition regime with a 55% chance, followed with a 27% probability of ending in the weak coupling regime state and only a 18% of ending in the strong coupling regime. Interestingly, once the Nb tip is functionalized with the TCNQ molecule, the coupling strength between the magnetic impurity and the STM tip can be modified in a reproducible and reversible manner (see section S7 in the supplementary material) by changing the tip sample distance. It has been proved before [33,34,50-53] that applying a mechanical force on the molecules on SC substrates, i.e decreasing the tip-sample distance, the interaction strength between the magnetic moment of molecules and the SC substrates can be modified. In our case, the molecule is attached to the SC tip and the forces applied to the molecule can be as well modified by changing the tip-sample distance, similarly as in atomic force spectroscopy experiments performed in Non-Contact AFM [54-56]. The relative tip-sample distance is controlled by applying a well-defined $\Delta Z$ offset after disabling the feedback loop and before the spectrum is recorded in order to observe the sub-gap states.

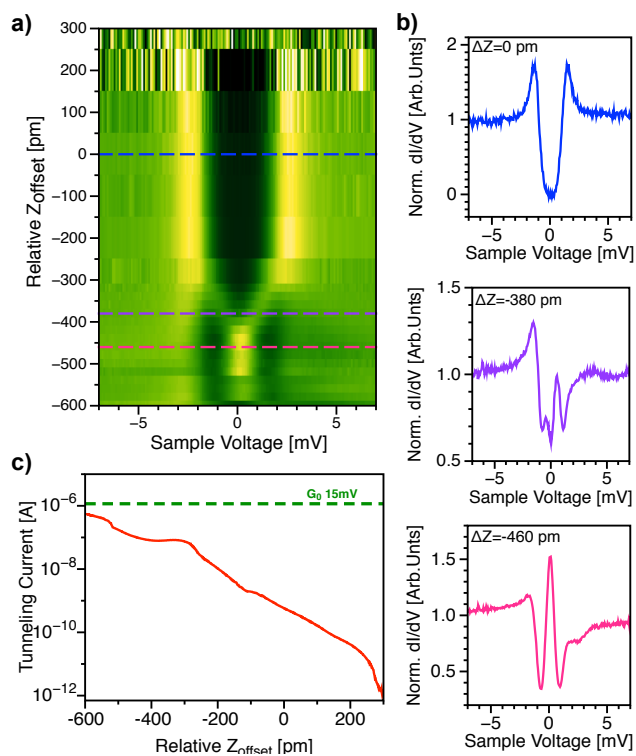

**Figure 3.** *Dependence of the STS spectra with the relative tip-sample distance. **a)** Normalized dI/dV intensity (green-yellow colour scale) as a function of bias voltage (horizontal axis) and relative tip-sample distance (vertical axis) for a TCNQ-functionalized tip and a gr/Ir(111) sample. The relative distance is controlled by applying a certain Z offset (negative values correspond to shorter distances) after opening the feedback loop prior to each STS acquisition, while maintaining the same stabilization parameters (V=15 mV, I=0.5 nA, $V_{mod}$ = 100 µV) throughout all the process; **b)** Three representative dI/dV spectra extracted from the stacked plot in a). **c)** I vs relative $Z_{offset}$ curve acquired with the same functionalized tip, covering the same 900 pm distance as in a). The dashed green line indicates the quantum of conductance, $G_0$. Stabilization parameters, V=15 mV, I=0.5 nA.*

Fig. 3(a) shows a representative example of the shift in energy of the sub-gap states as the functionalized SC Nb tip is moved closer to the gr/Ir sample. Upon decreasing the tip-sample distance, the sub-gap states emerge. As the tip approaches the sample, at -300 pm, the YSR states start to shift towards the Fermi energy (Fig. 3(a)), indicating an increase in the strength of the coupling between the molecular spin and the SC tip. The relative spectral weight of the bound states indicates that the impurity is initially in the weak coupling regime (see Fig. 3(b)). Both YSR peaks meet at zero bias for *relative $Z_{offset}$* = -460 pm and remain at this position for at least another 140 pm. The current versus *relative $Z_{offset}$* acquired with the same functionalized tip for this same distance range is shown in Fig. 3(c). A loss of the initial exponential dependence of the current coinciding with the emergence of the sub-gap states at -300 pm is observed. Although the quantum of conductance $G_0$ is never reached, this indicates that the forces between tip, molecule and substrate play a relevant role here. Note that the total change in distance is 900 pm. For comparison, in atomic force spectroscopy experiments, where the tip-sample distance is normally varied in the order of hundreds of pm, the frequency shift minimum is often crossed marking the change of force regime from attractive to repulsive [53-56]. We'd also like to highlight the fact that the process is completely reversible, as demonstrated in panels A and B of figure S7 in the supplementary material, where several reversibility tests are presented.

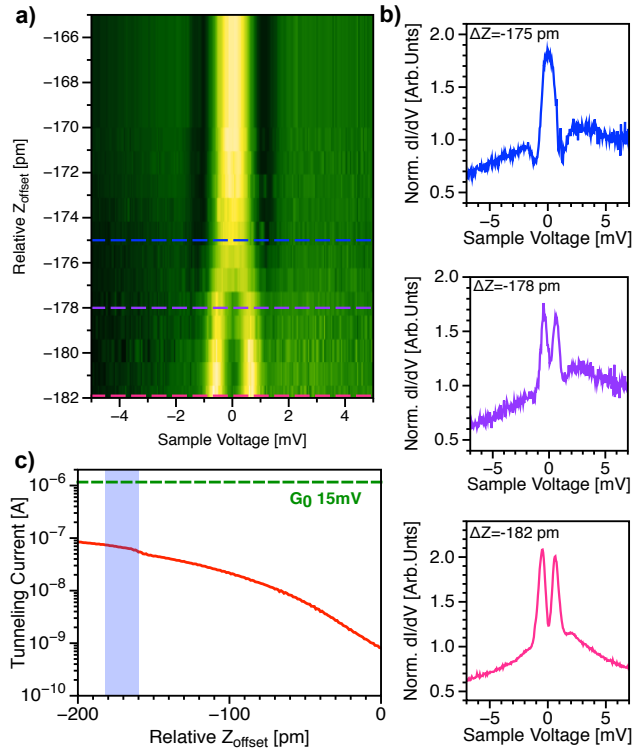

**Figure 4.** *Dependence of the STS spectra with the relative tip-sample distance. **a)** Normalized dI/dV intensity (green-yellow colour scale) as a function of bias voltage (horizontal axis) and relative tip-sample distance (vertical axis) for a TCNQ-functionalized tip and a gr/Ir(111) sample. In this case the transition is from a molecule in the quantum phase transition to a strongly coupled molecule. Stabilization parameters throughout all the process (V=15 mV, I=500 pA, V mod = 100 µV). **b)** Three representative dI/dV spectra extracted from the stacked plot in a). **c)** Long range I vs ΔZ curve acquired with the same functionalized tip, the 16 pm z-range shown in a) is highlighted in blue. The dashed green line indicates the quantum of conductance, $G_0$. Stabilization parameters, V=15 mV, I=500 pA.*

Fig 4(a) shows another experiment where the coupling regime of the magnetic impurity is modified by changing the tip-sample distance with a different TCNQ- functionalized Nb tip. In this case, when the tip-sample distance is reduced, the transition occurs from a molecule at the quantum phase transition to the strong coupling regime, as the coupling of the magnetic impurity to the superconductor increases. The tip is approached towards the sample a total distance of 200 pm (Fig. 4(c)), however, in the normalized dI/dV intensity maps only the section of the experiment where the bound states split is shown. The two bound states split at ΔZ= -176 pm and shift away from the Fermi energy for the next 6 pm. The relative spectral weight of the bound states indicates that the impurity ends up in the strong coupling regime. As in the previous example, the shift occurs well below the quantum of conductance, although a small bump can be observed in the corresponding region of the I(z) curve shown in panel c. The reversibility of this process can be checked in panels C and D of figure S7 in the SM. We'd also like to note that the same experiments have been carried out with clean, not functionalized, Nb tips as control experiments. The results shown in Fig. S3 of the supplementary material, demonstrate that no sub gap states appear throughout the experiment in a total length of 1010 pm. Finally, the possibility to control the interaction strength of a magnetic impurity manipulated onto a SC tip may open a new way to validate Matsuura's universal relation [27,28], as in principle the whole spectrum of the interaction strength of a magnetic molecule or adatom could be accessed by manipulating it onto a SC tip apex and performing spectroscopy experiments while changing the tip-sample distance (ΔZ) (see section S6 of SI for a more detailed explanation).

In conclusion, we show that by manipulating an initially uncharged TCNQ molecule from the surface of gr/Ir to the apex of a SC Nb tip the molecule charges and develops a magnetic moment. The exchange scattering potential produced by its spin results in the appearance of two symmetric sub-gap states. When an external magnetic field is applied to quench the superconductivity of the tip a Kondo resonance is recovered, demonstrating that the molecule has indeed developed a magnetic moment and that the sub-gap states observed are Yu-Shiba-Rushinov bound states [22,23]. The fact that the magnetic impurity is located in the SC tip apex may be used to tune continuously the exchange interaction, J, by controlling the distance between the functionalized-tip and the sample. In this work we have explored the transition between from the weak coupling regime to the QPS and from the QPS to the strong coupling regime. Also verifying the relation between the energy scale of YSR states, $E_{BS}$ and Kondo screening, $k_B T_K$, consistent in this case with the universal relation predicted by Matsuura [27,28] that connects the relative strength of both spin-dependent scattering processes.

This experimental approach could be used to enhance the sensitivity to the local spin texture by combining the Kondo [9] (to probe the magnetic exchange field) and YSR [10] (to probe spin polarization) regimes in the same probe, with the added value of being able to control the interaction strength between tip and molecule. A variety of magnetic systems can be investigated in this way, from adsorbed atoms and molecules to crystal surfaces, complementing other well-stablished spin polarized scanning probe techniques.

**Acknowledgements**
Work partially supported by the Ministerio de Ciencia, Innovación y Universidades projects PGC2018–093291–B–I00, PGC2018-097028–A–I00, PGC2018-098613–B–C21 and PID2019-109525RB-I00 and the Comunidad de Madrid (CM) projects Nanomag COST-CM,P2018/NMAT-4321 and NMAT2D-CM P2018/NMT- 4511. IMDEA Nanociencia acknowledges support from the 'Severo Ochoa' Programme for Centres of Excellence in R&D (MINECO, Grant SEV-2016-0686). IFIMAC acknowledges financial support from the Spanish Ministry of Science and Innovation, through the "María de Maeztu" Programme for Units of Excellence in R&D (CEX2018-000805-M. M.G. has received financial support through the Postdoctoral Junior Leader Fellowship Programme from "la Caixa" Banking Foundation (LCF/BQ/PI18/11630010).

## METHODS

**STM experiments**
All experiments have been carried out in a UHV chamber with a base pressure of $5 \times 10^{-11}$ mbar equipped with a Joule Thompson STM (JT-STM) and facilities for tip and sample preparation. The STM is fitted with a superconducting coil that can provide a magnetic field up to 3 T perpendicular to the plane of the sample and therefore parallel to the tip axis. All STM/STS data were measured with tip and sample thermalized at 1.2 K. In the case of our experimental setup the tunnelling voltage is applied always to the sample.

**Sample preparation**
The Ir(111) crystal was Ar+ sputtered and flash annealed in vacuum until a (1x1) LEED pattern and no traces of contaminants were observed. The graphene sample on Iridium were prepared via chemical vapour deposition (CVD) by exposing a clean Ir(111) crystal to a partial pressure of ethylene gas $C_2H_4$ ($P_{ethylene}$ = $5\ x\ 10^{-8}$ mbar) while the sample is held at T=1180 ºC for 10 minutes. TCNQ molecules are then thermally deposited from a quartz crucible heated to 80 ºC onto the clean surface of gr/Ir(111) held at room temperature. The coverage is kept below the monolayer by evaporating for less than 20 min. Subsequently, the sample was introduced into the STM and cooled to 1.2 K for inspection.

**Spectra Normalization**
All dI/dV spectra shown in the main manuscript and in the SI have been normalized in order to compare the different spectra. The STS spectra has been normalized by dividing the experimental data by the average value of the 10 first and last (total of 20 points) differential conductance values of the corresponding spectrum. In this way the edges of all spectra are always in the vicinity of the same dI/dV values.